\documentclass{article}

\usepackage[english]{babel}
\usepackage[utf8]{inputenc}
\usepackage[T1]{fontenc}
\usepackage{systeme,graphicx,multirow,comment,url, multicol}
\usepackage{hyperref, multicol, subcaption, color, authblk}
\usepackage[]{fullpage}

\newcommand{\bas}[1]{\footnote{\url{{#1}}}}
\newcommand{\email}[1]{\href{mailto:#1}{\nolinkurl{#1}}} 

\newcommand{\etal}{\emph{et al.~}}
\newcommand{\ie}{\emph{i.e.~}}

\begin{document}

\title{Multi-scale simulation of COVID-19 epidemics\footnote{This work has been published and presented at the international GAMA days held in June 2021 \cite{gamadays21}.}\footnote{This internship and report have been validated for completion of the ENSIMAG engineering degree in 2021.}}
\author[1]{Benoit Doussin}
\author[2]{Carole Adam}
\author[3]{Didier Georges}
\affil[1]{ENSIMAG, Grenoble INP, benoit.doussin@grenoble-inp.org}
\affil[2]{Univ. Grenoble-Alpes, LIG, carole.adam@imag.fr}
\affil[3]{GIPSA-lab, Grenoble-INP, didier.georges@gipsa-lab.grenoble-inp.fr}
\date{\textbf{\emph{This document is based on the M2 internship report of Benoit Doussin for the\\ ENSIMAG engineer degree, conducted between February and \\June 2021 under the supervision of Carole Adam and Didier Georges}}}
\maketitle

\paragraph{Abstract}
Over a year after the start of the COVID-19 epidemics, we are still facing the virus and it is hard to correctly predict its future spread over weeks to come, as well as the impacts of potential political interventions. Current epidemic models mainly fall in two approaches: \textbf{compartmental models} divide the population in epidemiological classes and rely on the mathematical resolution of differential equations to give a \textbf{macroscopic} view of the epidemical dynamics, allowing to evaluate its spread a posteriori; \textbf{agent-based models} are computer models that give a \textbf{microscopic} view of the situation, since each human is modelled as one autonomous agent, allowing to study the epidemical dynamics in relation to (heterogeneous) individual behaviours. 
In this work, we compared both methodologies and combined them to try and take advantage of the benefits of each, and to overcome their limits. In particular, agent-based simulation can be used to refine the values of the parameters of a compartmental model, or to predict how these values evolve depending on sanitary policies applied. In this report we discuss the conditions of such a combination of approaches, and future improvements.

\section{Introduction}

It has been over a year since the COVID-19 epidemic began, and we are still facing virus. Public authorities have adopted several strategies, differing between countries, in order to protect us. Before massive screening was available, in order to reduce the increase in the number of hospitalisations and the overcrowding of intensive care units, public authorities had to set a first very strict lockdown. This strict lockdown proved to be particularly effective, and it allowed us to subsequently spend an almost "normal" summer vacation. Further prevention policies have then been decided, such as the requirement to wear a face mask in public places, or the increase in screening tests. Among these tests, we can distinguish: \textbf{virological and antigenic tests}, which determine whether the person is currently carrying the disease; and \textbf{serological tests}, which detect the presence of antibodies in order to determine whether the person has ever been a carrier of the disease.

Today it is difficult to make even a provisional assessment. How many people have already been infected, or what has been the real impact of the health policies put in place to fight the disease? The aim of this report is to try to propose tools that could help to provide some answers. Concretely, we study how the combination of two different approaches allows to determine the impact of a new health policy, and to predict the evolution of the epidemic as accurately as possible.

\section{Background}

Models concern the notion of resemblance as well as imitation. They are meant to represent reality, even if they cannot be perfectly similar. Models can be used to predict, to test, to learn or to help in decision making. We will focus here on two types of models, namely SIR models and agent-based models.

\subsection{Compartmental models}

Compartmental models divide the population in a number of different compartments or classes, corresponding to different epidemiological statuses (susceptible, infected, recovered...). Differential equations then express the evolution of the number of people in each compartment over time. Such models provide an aggregated, macroscopic view of the epidemical dynamics.

\subsubsection{SIRD model}
The SIRD model, which is the simplest, allows us to see how the variables fit together. The population is divided into four compartments: Susceptible, Infected, Recovered, Dead, as follows:
\begin{multicols}{2}
\begin{center}
\includegraphics[scale=0.2]{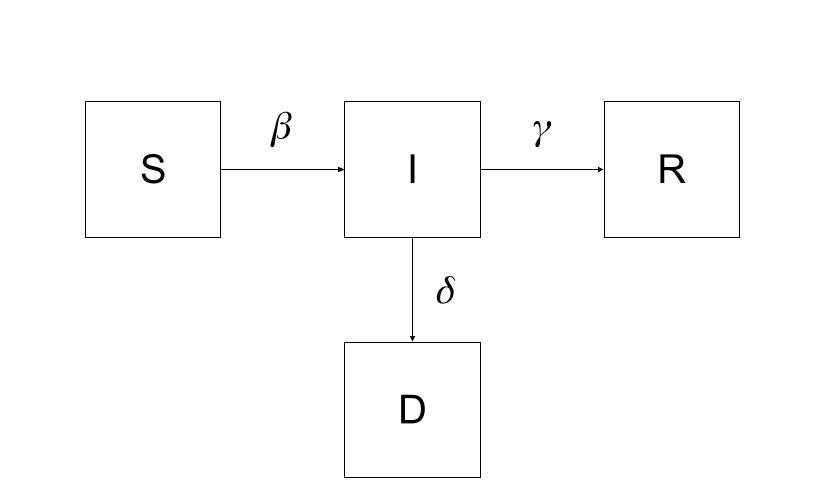}
\end{center}
\strut \vspace*{-7pt}
\begin{center}
$\systeme{
\frac{dS(t)}{dt} = -\beta (t) I(t)S(t),
\frac{dI(t)}{dt} = \beta (t) I(t)S(t) - (\gamma + \delta)I(t),
\frac{dR(t)}{dt} = \delta I(t),
\frac{dD(t)}{dt} = \gamma I(t)
}$
\end{center}
\end{multicols}

We have, at a time $t$, a certain number of people in each compartment. We then consider that at time $t+1$, the difference will depend on the number of people in the compartments as well as on various variables. Here $\beta(t)I(t)$ is called the force of infection, and shows that the number of healthy individuals decreases proportionally to the number of infected individuals. In this model, we can consider $\delta$ and $\gamma$ as constants. On the other hand, the variable $\beta$ might evolve over time, because it can be impacted by several parameters, such as the policy set up by the country to fight the epidemic, or just the media coverage of the disease which could incite people to be more careful.

It is also interesting to know that it is on the basis of such SIRD models that the basic reproduction number, also known as $R_{0}$ and that got huge media cover, is computed. This $R_{0}$ represents the average number of people that one individual will infect while being sick. If $R_{0} < 1$, then it means that $\frac{dI(t)}{dt} < 0$, so the number of infected people decreases over time. We then seek to have:
\begin{center}
    $
    \frac{\beta(t)S(t)}{\gamma + \delta} < 1
    $
\end{center}
If a government wants the virus to disappear from the population, they must then manage to maintain $R_{0} < 1$, and thus $\beta(t) < \frac{\gamma + \delta}{S(t)}$. On the other hand, if $R_{0} > 1$, then $\frac{dI(t)}{dt} > 0$, which means that the number of infected people will keep increasing over time. We can generalise the value of $R_{0}(t)$ with:
\begin{center}
    $
    R_{0}(t) = \frac{\beta(t)S(t)}{\gamma + \delta}
    $
\end{center}

This model thus provides interesting insight, but in the context of SARS-Cov-2, it is not sufficient. Indeed, it supposes that the real number of infected people is known. Yet, given that the proportion of asymptomatic people is unknown, and that the number of screening tests was very low at the beginning of the epidemic, this real number of infected people is not available and can only be approximated. Therefore, we need to find a more relevant model.

\subsubsection{Epidemic model with non linear infection force}

Wang \cite{wang2006epidemic} has an interesting approach because it addresses the case of SIR models with a non-linear infection force. The model described presents the force of infection as a function of $I(t)$, noted $\beta(I(t))$. Without any intervention to stop the virus from spreading, this infection force could be considered as $\beta I(t)$, where $\beta$ is a constant. 
Wang then studies the SIR model by changing the form of the infection force, and in particular in the case of a SIR model with an intervention policy. He therefore factors his function $\beta(I(t))$ by $\frac{\lambda I(t)}{f(I(t))}$, where $f(I(t))$ is the impact of the policy on the transmission coefficient $\lambda$. He succeeds in showing that for the model with instantiation of sanitary policies to fight the virus, even if the force of infection is non-linear, there is a disease-free equilibrium, if the reproduction number $R_{0} < 1$.

This article is interesting because it addresses the case of political intervention, which echoes the current crisis and the measures put in place to manage it. It shows that it is possible to contain the spread of the virus. However, the model remains incomplete as it does not take into account the distinction between the symptomatic and the asymptomatic, which is however important because the symptomatic compartment has no impact on the spread (it could be considered as a quarantine compartment). It is also difficult to simplify $\beta(t)$ by $\frac{\lambda}{f(I(t))}$ because various different policies have been implemented over the last year, and therefore the function $f$ has not always been the same.

\subsubsection{Charpentier's model}

In the context of the COVID-19 crisis, a simple SIR model is not sufficient to correctly model the spread of the virus in France. An interesting model has been proposed by Charpentier \etal \cite{charpentier2020covid} to best describe the current crisis. This model is much more complete because it takes into account a greater number of compartments, in particular the undetected infected $I^{-}$ (who spread the virus silently), as well as the hospitalised $H$ and the people in resuscitation $U$ (who put pressure on the care system). These last two compartments are particularly important, since fairly reliable data is available, which will allow a more relevant and precise fitting of the model using regression methods.

\begin{multicols}{2}

\begin{center}
\includegraphics[scale=0.2]{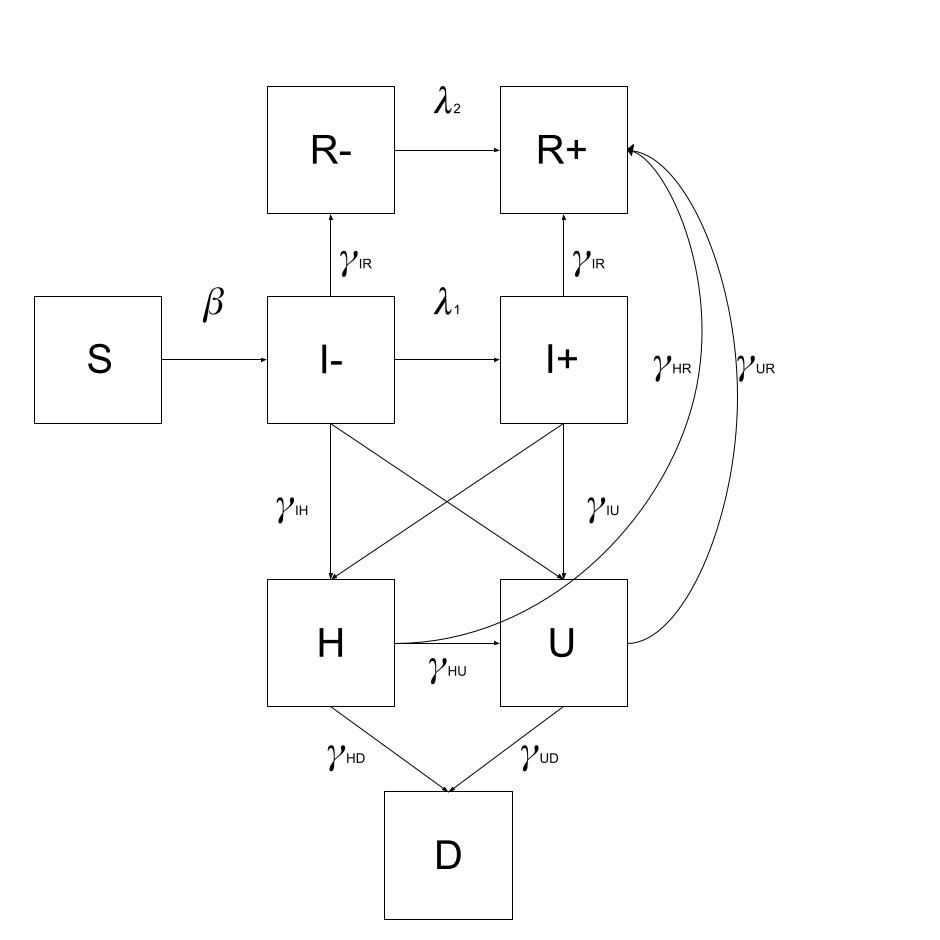}
\end{center}

\begin{center}
$\systeme{
\frac{dS(t)}{dt} = -\beta (t) I^{-}(t)S(t),
\frac{dI^{-}(t)}{dt} = \beta (t) I^{-}(t)S(t) - \lambda_{1} I^{-} - (\gamma_{IH} + \gamma_{IU} + \gamma_{IR})I^{-},
\frac{dI^{+}(t)}{dt} = \lambda_{1} I^{-} - (\gamma_{IH} + \gamma_{IU} + \gamma_{IR})I^{+},
\frac{dR^{-}(t)}{dt} = \gamma_{IR}I^{-} - \lambda_{2}R^{-},
\frac{dR^{+}(t)}{dt} = \gamma_{IR}I^{+} + \gamma_{HR}H + \gamma_{UR}U + \lambda_{2}R^{-},
\frac{dH(t)}{dt} = \gamma_{IH}(I^{-}+I^{+}) - (\gamma_{HU} + \gamma_{HD} + \gamma_{HR})H,
\frac{dU(t)}{dt} = \gamma_{IU}(I^{-}+I^{+}) + \gamma_{HU}H - (\gamma_{UR} + \gamma_{UD})U,
\frac{dD(t)}{dt} = \gamma_{UD}U + \gamma_{HD}D
}$
\end{center}
\end{multicols}

In this model, $\lambda_{1}$ represents the rate of people found to be positive by virological or antigenic tests, and $\lambda_{2}$ represents the rate of people found to have been in contact with the disease by serological tests.

The $\gamma$ are the transition rates between the compartments. We can consider that they remain constant, even if we can intuit that this is not really the case. For example $\gamma_{UD}$, in the event of saturation of the hospitals, would risk increasing considerably because of the limited number of beds and doctors, when many people transition from $U$ (ressuscitation) to $D$ (dead). 
\enlargethispage{20pt}
Didier Georges \etal \cite{guan2020transport} study the spread of SARS-COV-2 in France using the Charpentier model. The work done in the Simulation section of this report is also based on their work.

\subsection{Agent-based models}

Multi-agent models are computer models composed of a set of autonomous agents located in an environment and interacting according to certain relationships and situations. Each agent has its own decision system and will act accordingly: theese models are at a microscopic scale. They make it possible to simulate particular social situations and to see how a population will react to them, at the macroscopic scale. For example, how will the virus spread if some pupils in a school do not respect the distances between them? Or how can the risk of infection increase during rush hour in public transport?

The main benefits of this kind of simulation are, on the one hand, that it is particularly visual, allowing to watch the interactions between the agents. On the other hand, agent-based models allow to simulate heterogeneous individual human behaviour, at the microscopic level, where compartmental models act like black boxes, they only provide the macroscopic or aggregated view (proportion of people in each compartment), but does not show how each individual acts and how it relates to their epidemiological status. For instance, agent-based models allow to investigate questions such as: are people not wearing masks more often infected?

\subsubsection{Epidemiological Agent-Based Model for Irish Towns}
Hunter \etal propose an agent-based model of Irish cities \cite{hunter2016open, hunter2018comparison}, implemented in Netlogo \cite{netlogo}. They are also looking to draw parallels between SIR models and agent-based simulations. They show that both approaches can indeed give similar results, but that agent-based models capture the stochasticity that exists in the world. On the other hand, SIR models have a much lower computational cost, especially when trying to calculate the propagation over a large population. However, as their model was implemented before the appearance of SARS-COV-2, they did not take into account asymptomatic people, which is an important factor in the spread of the virus today and must therefore be taken into account.

\subsubsection{Models for decision support}

\begin{figure}
\begin{center}
\includegraphics[scale=0.30]{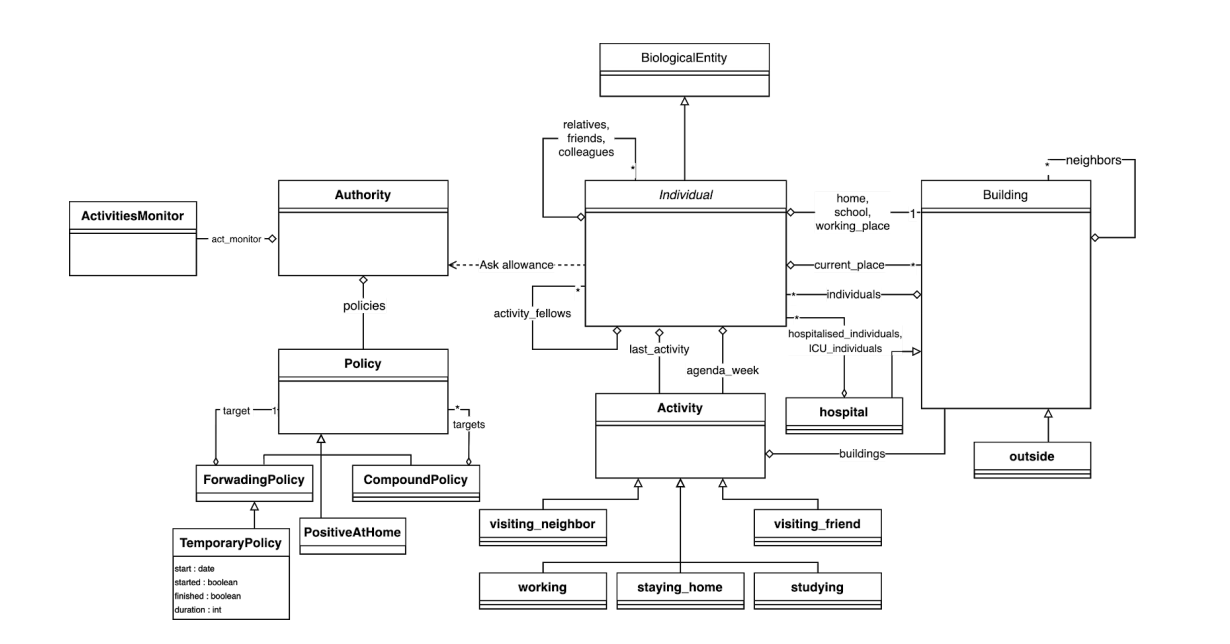}
\caption{UML of COMOKIT model, from \cite{brugiere2010odd}} \label{uml:comokit}
\end{center}
\end{figure}

COMOKIT \cite{gaudou2020comokit} is a project originally developed to meet the needs of the Vietnamese government and assist them in decision making. It has been coded in GAMA, and one of its strengths is that it is modular and scalable. The data can be changed to study other cities, which was done to study the spread of the virus in Nice \cite{chapuis2021using}. It was designed to assist decision making and has really been used for this, in particular in Vietnam. Figure~\ref{uml:comokit} shows the UML diagram of the COMOKIT model, summarising the different agents included.
COMOKIT allows to explore the impact of the different policies put in place. The model is very comprehensive. It takes into account a large number of parameters, which enables it to use a particularly realistic synthetic population. Another interesting point is that it takes into account the fact that an agent can also fall ill without coming across anyone, \ie simply by entering a building where the ambient air is contaminated. On the other hand, this model assumes that a person will not get sick outside, especially when travelling, but only in enclosed spaces. By making this assumption, they can afford to use a time step of 1 hour to improve computational performance, but as a result they fail to capture the possible contaminations in public transport.

Another multi-agent model that can be cited is the one proposed by Neil Fergusson \cite{ferguson2020report}, that has been instrumental in the decision to set a lockdown in England. This model is based on previous work done by the team on other viruses, that was adapted to COVID-19 by tweaking the parameters.

Pascarmona \etal \cite{pescarmona2020can} present a multi-agent model with a SISaR approach, \ie with differentiated compartments for symptomatic/asymptomatic infected people. Their model is implemented in Netlogo\footnote{SISaR model available at: \url{https://terna.to.it/simul/SIsaR.html}}. The authors therefore consider four compartments: Susceptible, Symptomatic, Asymptomatic, and Recovered. They use their simulator to compare various scenarios with different policies, providing interesting insight. For example, they show that opening schools has little impact on the virus spread. However, they do not consider Hospitalized or Intensive Care, which we need to fit our model to available data. 

\subsubsection{Explanatory models}

The Covprehension consortium \footnote{CovPrehension website and simulator: \url{https://covprehension.org/simulator/}} has designed a website that provides scientific popularisation articles accompanied by interactive epidemics simulators. Their goal is to answer questions from the general public about the COVID-19 epidemics, with interactive simulators\cite{covprehension-rofass} that people can play with to test and understand the impact of sanitary policies on the virus spread \cite{cnrs-simu-1, cnrs-simu-2}.

These models and simulations are deliberately simple \cite{axelrod1997advancing}, with the goal of letting people visualise the dynamics of the epidemics and understand its mechanisms. The objective is education and prevention, rather than prediction. As such, the model does not require a high degree of realism nor a great complexity \cite{axelrod1997complexity}. However, it is interesting to note that despite the simplicity of the underlying models, the epidemics curves produced are almost similar to those produced by SIR models.

\subsubsection{Agent-Based Social Simulation of the Covid-19 Pandemic}

A great number of models exist that cannot all be presented here. Lorig \etal \cite{lorig2021} realise a survey of 126 published articles that use agent-based modelling and simulation of the COVID-19 pandemics. 
This survey is very interesting to situate the current model with respect to the literature. It is for instance surprising that only 50\% of the models analysed in this survey make a distinction between symptomatic and asymptomatic infected people. Such a distinction might not have been useful for other types of viruses, but in the case of SARS-COV-2, asymptomatic contagion plays a major role in the sperad of the virus. Indeed, people who are sick and aware that they are sick will most probably stay in confinement (wether willingly or forced by sanitary rules), therefore protecting others, while asymptomatic people have no reason to change their behaviour, and will therefore silently spread the virus. A model must make this distinction to precisely capture the dynamics of the COVID-19 epidemics. 

This survey also offers a critical viewpoint on existing models and gives an idea of required improvements in these models. For instance, important parameters that are not always taken into account in reviewed models include the fact of wearing a mask or of respecting physical distanciation.

\subsection{Synthesis of the approaches}

\subsubsection{Comparing the 2 approaches}
Both approaches, although they may be intended to simulate the same thing, have their advantages and disadvantages due to their different methodologies. In particular, each view can be either beneficial or unfavourable depending on the context and goal of the model. 

One criticism of compartmental models is that they act like black boxes. Indeed, even if the aggregated result obtained is correct, some information is lacking: are people who take public transport more frequently infected? Does working in a city centre increase the risk of contracting the virus? This is because SIR models only give an aggregated view of the number of individuals in each compartment, but without indicating who these individuals are. On the other hand, agent-based models do provide this information, but they cannot be used for large-scale simulations because they are too computationally intensive. Being simpler to compute, SIR models can be used at the scale of a city, a region or an entire country, offering interesting insight into the global spread of the epidemics.

Another notable advantage of agent-based models is their ability to account for the heterogeneity of the population. For example, such a model could include individual attributes representing weariness setting in after months of restrictions, or a tendency to disobey rules. These aspects would be captured in a SIR model at the aggregated level only, translating into an increase in the transmission rate $\beta$, which is less visual and less explainable. Agent-based models therefore seem more adapted for education or popularisation goals.

\begin{center}
    \begin{tabular}{|c|c|c|}
    \hline
         & Compartmental models & Agent-Based models \\
    \hline
        \multirow{2}{*}{Benefits} & Multi-scale & Microscopic view of individual behaviour \\
        & Computationally light, fast & Heterogeneous population \\
    \hline
         \multirow{2}{*}{Drawbacks} & A posteriori analysis & Limited to small scale \\
         & No information about individuals & High computational cost\\
    \hline
    \end{tabular}
\end{center}

\subsubsection{Combining the approaches}

In the articles studied, multi-agent models often seek to simulate SIR compartments, \ie to consider that in addition to its own characteristics, each agent has an epidemiological state that classifies it in one of the compartments. This is the most relevant method because it allows the dynamics of SARS-COV-2 to be described very finely, while retaining demographic stochasticity. 

However, when an article refers to both the multi-agent model and the compartmental model, it is usually to make comparisons between the two approaches. Also, multi-agent simulations are often used to determine the best policy strategy to implement.

\cite{colizza2016} also notes that no model currently exists that combines the scalability of compartmental models to the global level, with the fine-grained description of the population with individual specificities. They state as a future project the integration of agent-based and compartmental models into GLEAM-France, a more comprehensive model intended at informing the public health response strategy in France.

In this study, we also seek to combine the two approaches, namely compartmental and agent-based paradigms. Concretely, instead of making simple comparisons, is it possible to extrapolate the results obtained using multi-agent models to determine the new parameters of a compartmental model? In this way, the future propagation on a departmental scale could be determined without the computational cost that would come with a million-agent model.

\section{Our proposed model}

Our model is described following the ODD protocol \cite{oddProtocol} and its overview section, that gives a clear idea of the implemented model.

\subsection{Objective}
The objective of our model is to study if agent-based models can be combined with compartmental models to improve the simulation of the dynamics of the spread of SARS-COV-2. We focus on the city of Lumbin (Isère, France) during the second lockdown in France (from 30 October to 15 December 2020). By introducing various sanitary constraints in the agent-based model, we seek to observe how the virus propagation changes as a result, and measure the new values of parameters (transmission rate, etc) to be fed into the compartmental model.

\subsection{Agents, Variables, Scale}
The model aims to study the propagation of the virus on a microscopic scale, which is why each individual will be represented by a $Habitant$ agent with its own characteristics, and in particular its health state. Each day, an individual will perform a certain number of tasks according to his schedule, which will have been generated beforehand.

The $Habitant$ agents are distributed in 3 age groups: children, active adults, and retired people. Each resident of the city of Lumbin has an individual agenda that lists the dates and times when they must realise specific activities. Possible activities include: $Rest$, $Work$, $Shop$ et $GoToPub$. Each of these activities is associated with a place, represented by a spatial agent called $Building$. Agents also have a group of friends with whom they can socialise. Residents all have a health status among Susceptible, Infected (Symptomatic or Asymptomatic), or Recovered. 

Symptomatic infected people are therefore aware that they are sick, and can choose to stay confined at home, if they respect sanitary measures. However, we consider that not all residents do systematically respect these measures (social distancing, isolation when sick, wearing a mask...). Respecting those constraints decreases the probability to get infected and to infect others, so agents not respecting them have more risks to get sick and to spread the virus. We also consider that while at home, people to not maintain any sanitary precautions with their family. 

$Building$ are distributed into 5 different categories as follows: 
\begin{itemize}
\item $Residential$ are where the $Habitant$ agents live.
\item $Industrial$ represent office buildings, where a part of the active population works.
\item $Educational$ are where children spend the day studying.
\item $Commercial$ are shops where adults go for shopping, and also where a part of the active population works.
\item $Restauration$ where $Habitant$ agents can socialise with their friends around drinks or food.
\end{itemize}
It must be noted that during the second lockdown in France, the $Restaurantion$ buildings (pubs and restaurants) were closed. As a result, it was not possible for people to go to these places, neither to socialise, nor to work; people usually working in pubs and restaurants are considered to work from home.

\begin{figure}[hbt!]
    \centering
    \includegraphics[scale=0.37]{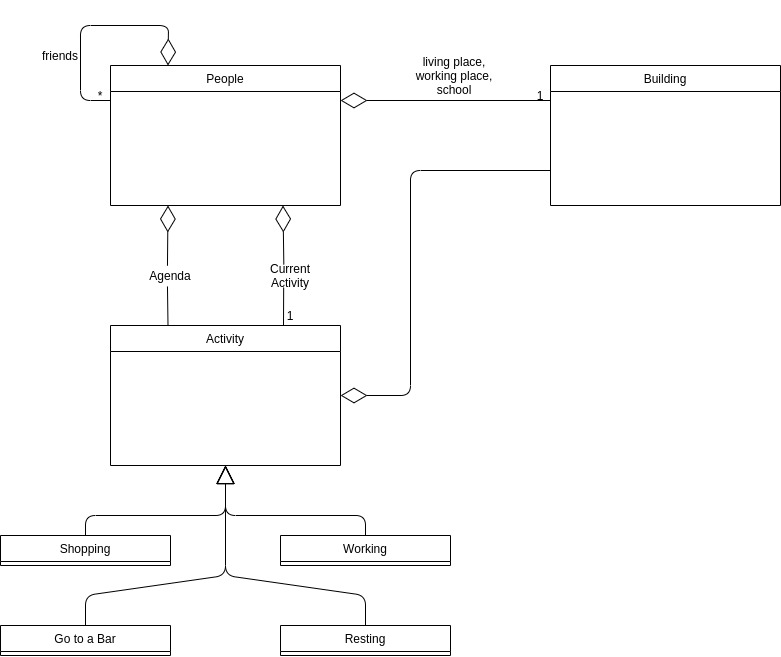}
    \caption{UML description of our model} \label{fig:uml}
\end{figure}

All these agents are shown on the UML diagram of our model in Figure~\ref{fig:uml}.

\subsection{Global view of the process, planning}
The model represents the daily life of residents of Lumbin during the lockdown, with a time step of 10 minutes. At each step, a resident will perform an action depending on current time, agenda, and health state. When it is time to work, active residents who cannot work from home will head to their work place. Those who can work from home, or those who are sick with symptoms, will stay home. When work time is over, active residents head back home. The same applies to children but they go to school instead, and they cannot study from home; they only stay home when sick with symptoms. Active residents and retired residents can also go out to shop during their free time, on condition that they respect the curfew imposed by the sanitary policies in place. 

At each time step, sick residents can recover, but can also infect susceptible residents around them, with a certain probability. This infection probability also depends on the residents' behaviours, and increases for people not respecting the protective gestures: they have more risks to get infected, and to infect others.

\section{Simulation}

\subsection{Study on Charpentier's model}
For the Charpentier model, we are interested here in Isère, for which we know the number of people hospitalised (H), in intensive care units (U), and deaths (D) per day. This data is quite reliable and will allow us to adjust the model in order to obtain a realistic trend of the $\beta(t)$ parameter. One of the approximations carried out is to consider $\lambda_{2} = 0$, because the serological tests are not really put forward and especially do not play any role on the value of $\beta(t)$. This first approximation has little impact on the calculations because we are mainly looking for the current number of susceptible people, and we adjust the model on the reliable data available about the number of H, U and D observed\footnote{Open Data for coronavirus in Isère: \url{https://opendata.isere.fr/explore/?sort=modified&refine.theme=Covid19}}.

In order to determine the $\beta$ coefficients, a sliding horizon approach must be used, and therefore restricted to a window $[t, t+k]$. We must then find the parameters which will allow the best curve fit by minimizing the residual. For this we need an iterative method: we will use the Gauss-Newton algorithm, which allows a non-linear least squares resolution. What we are seeking are thus the values of parameters $\alpha$ such that :
\begin{center}
    $min \sum_{i=t}^{t+k} (Z_{fit}(i, \alpha) - Z_{obs}(i))^{2}$
\end{center}
with $Z_{fit}(i, \alpha)$ the output vector predicted for parameters $\alpha$, and $Z_{obs}(k)$ the observed vector. To realise the parameterisation, we rely here on the number of H, U and D observed in Isère, by setting parameters $\gamma$ and varying $\beta$ and $\lambda_{1}$.

\begin{figure}[hbt!]
\begin{subfigure}{0.5\textwidth}
    \centering
    \includegraphics[scale=0.2]{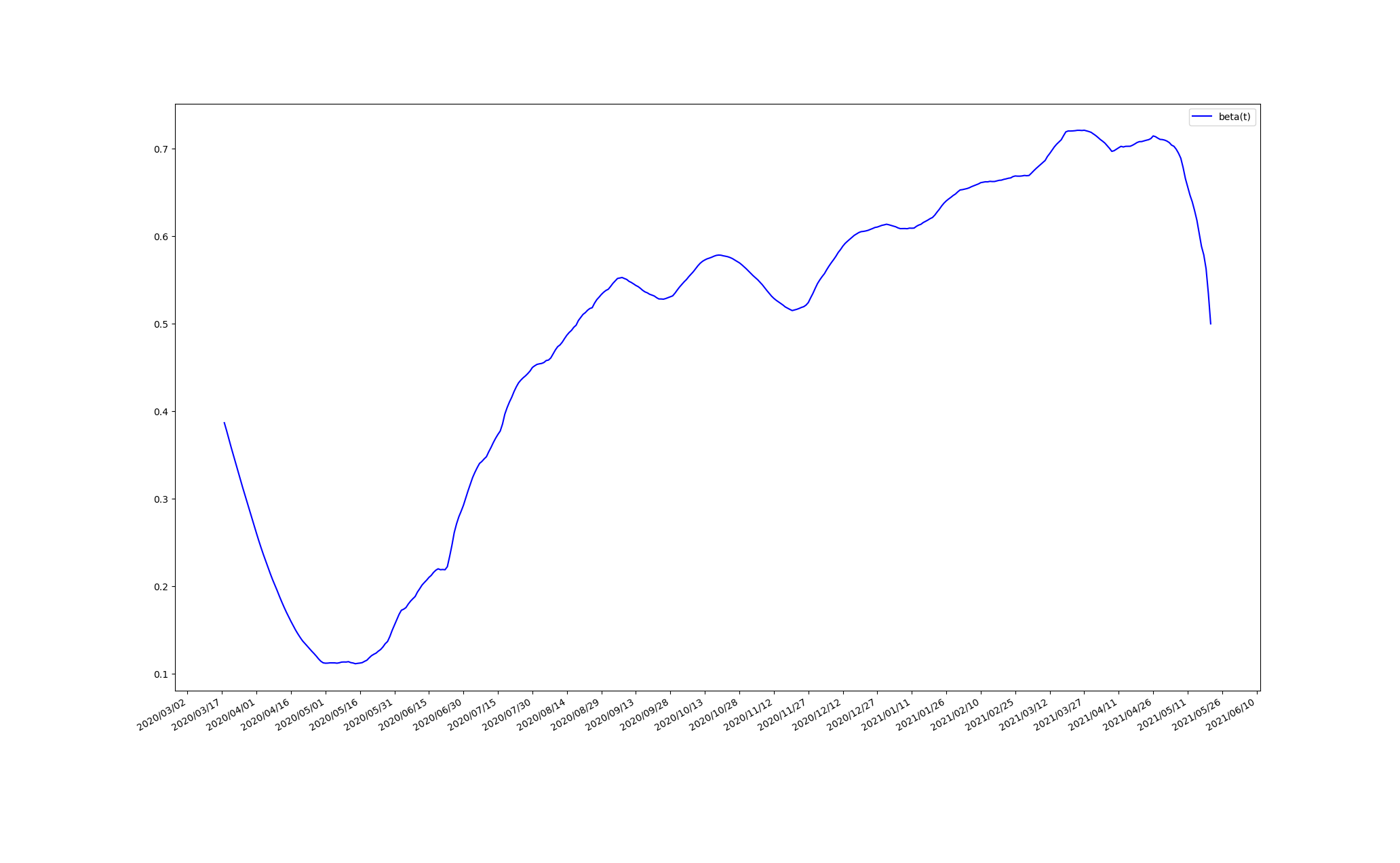}
    \caption{$\beta(t)$ estimated} \label{fig:beta}
\end{subfigure}
\begin{subfigure}{0.5\textwidth}
    \centering
    \includegraphics[scale=0.2]{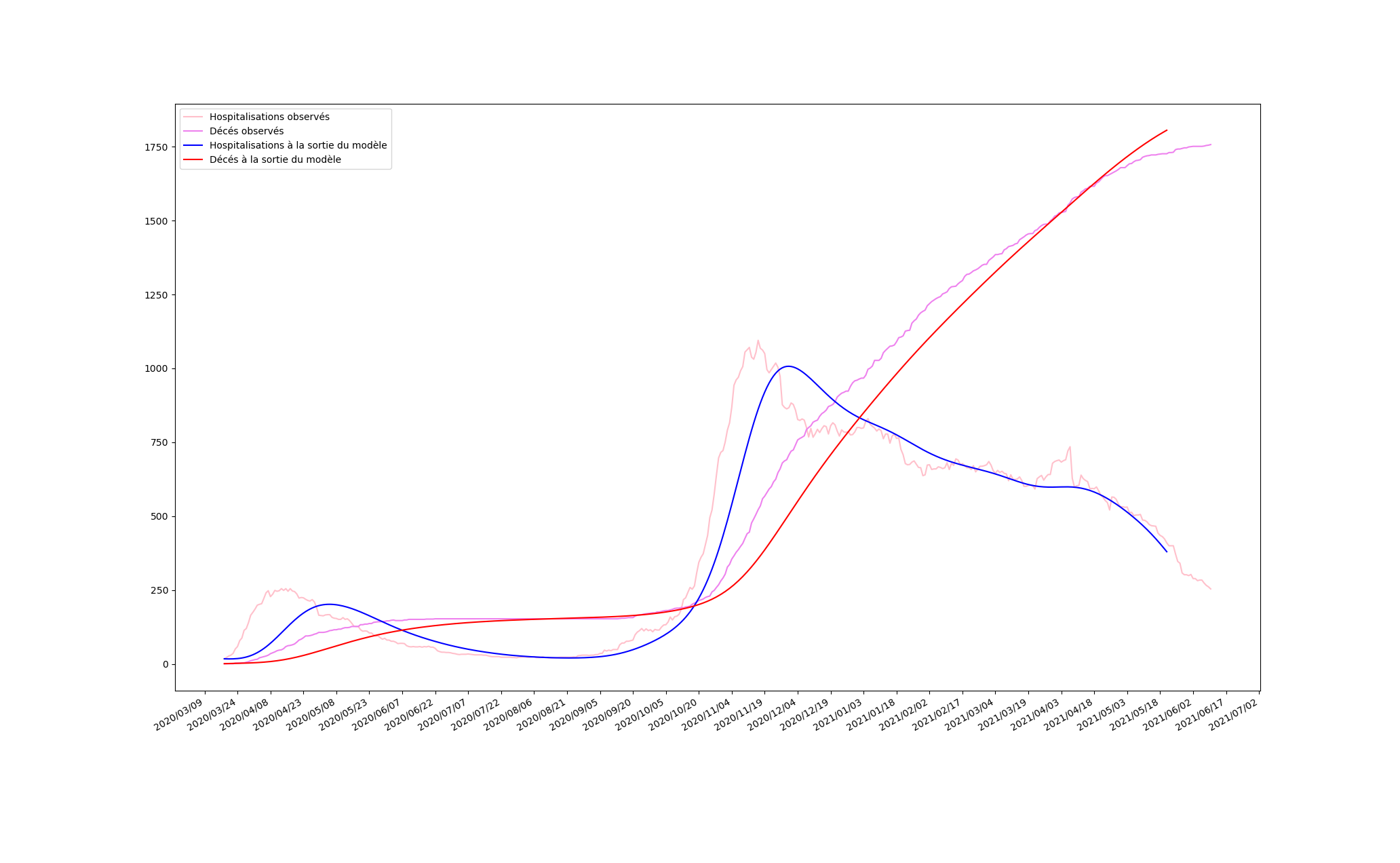}
    \caption{$H(t)$ and $U(t)$ observed vs $H(t)$ and $U(t)$ estimated as output of the model} \label{fig:hu}
\end{subfigure}
\end{figure}

\begin{figure}[hbt!]
    \centering
    \includegraphics[scale=0.2]{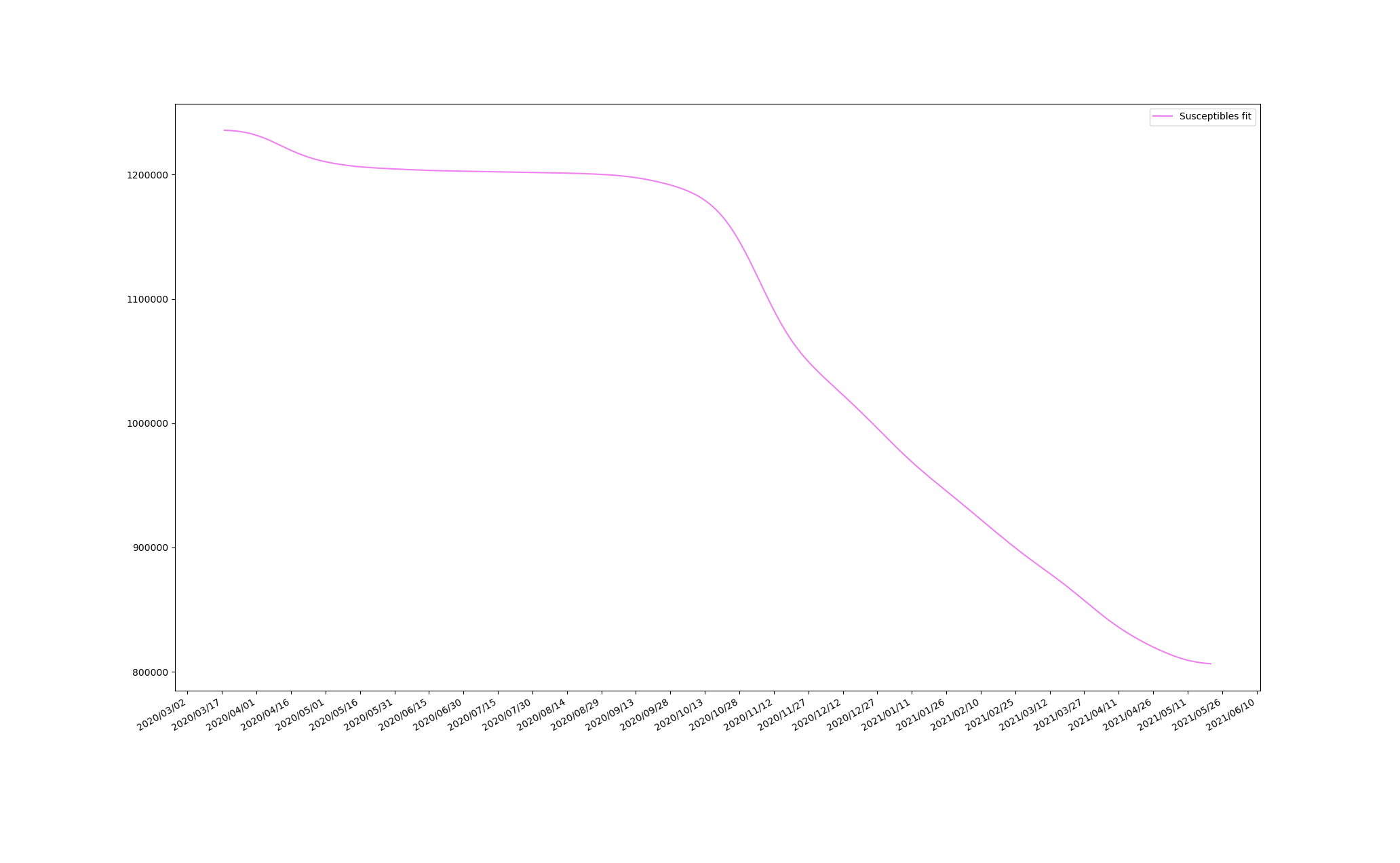}
    \caption{$S(t)$ estimated} \label{fig:st}
\end{figure}

Using the method described above, we obtain the results shown in Figures~\ref{fig:beta}, \ref{fig:hu} and \ref{fig:st}. The value of $\beta(t)$ (Figure~\ref{fig:beta}) is probably over-estimated; however, it gives a concrete idea of the influence of sanitary policies on its evolution. The time $t=0$ is the date of 17 March 2020, the start of the first lockdown: the graph clearly shows a strong impact of this strict lockdown on the value of $\beta$. Future experiments should play on the values of parameters $\gamma$ to obtain a more precise model.

By looking at the trend of the estimated $S(t)$ (Figure~\ref{fig:st}) we can see that in Isère, approximately 30\% of the population has already been infected by SARS-Cov-2. This estimation is close to the computations of Institut Pasteur\footnote{Proportion de la population ayant été infectée par SARS-CoV-2 : \url{https://modelisation-covid19.pasteur.fr/realtime-analysis/infected-population/}}, that estimates that 25\% of the population of the Auvergne-Rhône-Alpes region (that includes the Isère department) has already been infected.

\subsection{Agent-based model}

\subsubsection{Implementation}

\begin{figure}[hbt!]
    \centering
    \includegraphics[scale=0.25]{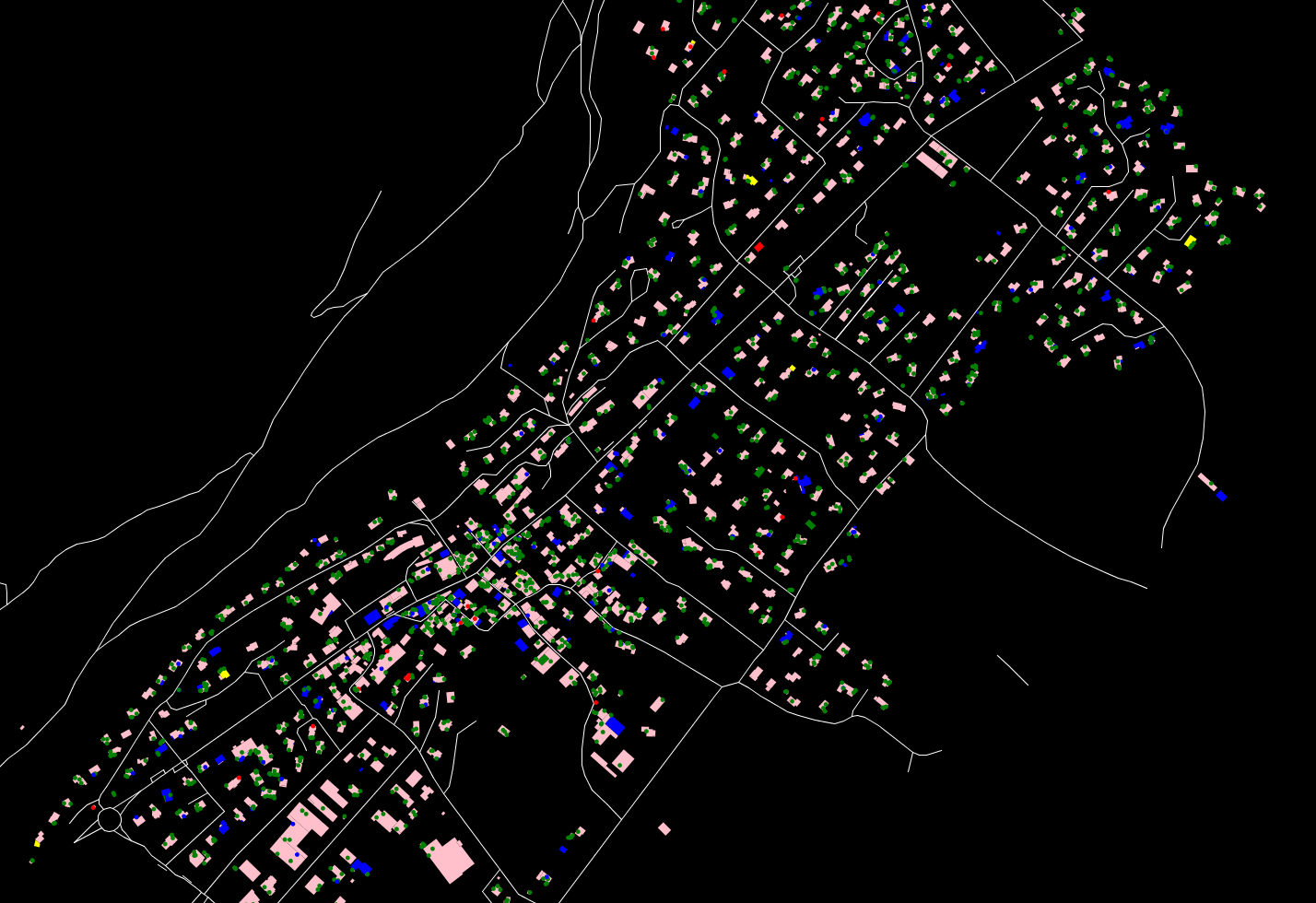}
    \caption{City of Lumbin as simulated in GAMA} \label{fig:GamaL}
\end{figure}

The model presented above was implemented in GAMA\cite{taillandier2010gama}, an agent-based simulation platform that facilitates the integration of GIS data. The model is applied on the city of Lumbin, with 2143 residents according to the 2015 census data. This number is small enough to allow the initialisation and simulation of each individual resident. Geographical data about the town was recovered as a shapefile with the QGIS software, and integrated in the model. This provides a more realistic model of the town, and a more realistic simulation of the population movements. The simulation is also more visual, as can be seen on the screenshot in Figure~\ref{fig:GamaL}. This screenshot of our simulator shows residential buildings in pink, and industrial buildings in blue.

\subsubsection{Initialisation}

Since a heterogeneous population within a city is being studied, several parameters are drawn randomly, while ensuring that their mean matches the observations. At initialisation, some attributes are randomly generated for each individual: \enlargethispage{10pt}
    \begin{itemize}
        \item Their age group;
        \item The location of their residence and workplace (or school);
        \item Their agenda (time at which they realise their activities, in particular time when they start and finish work, and if they prefer shopping in the morning or afternoon);
        \item The places where they socialise;
        \item Whether they can work from home (except from people working in shops and schools);
        \item Acquaintances: work colleagues; people living in the same building; friends with whom they socialise.
        \item Whether they respect social distancing in public spaces.
    \end{itemize}

\noindent Some parameters of the virus and its spread are also drawn randomly around the observed mean:
    \begin{itemize}
        \item The exact duration during which an agent stays infected;
        \item The probability to have symptoms or not;
        \item The probability to be infected when in contact with an infected agent.
    \end{itemize}

\subsubsection{Calibration}

Due to this randomness, running the same simulation twice might give very different results. A first part of this work was therefore dedicated to studying the convergence of the average in order to determine the number of iterations necessary for the model to converge to a stable average. It appears on Figure~ref{fig:convmoy} that the average stabilises after about thirty simulations. 

\begin{figure}[hbt!]
    \centering
    \includegraphics[scale=0.2]{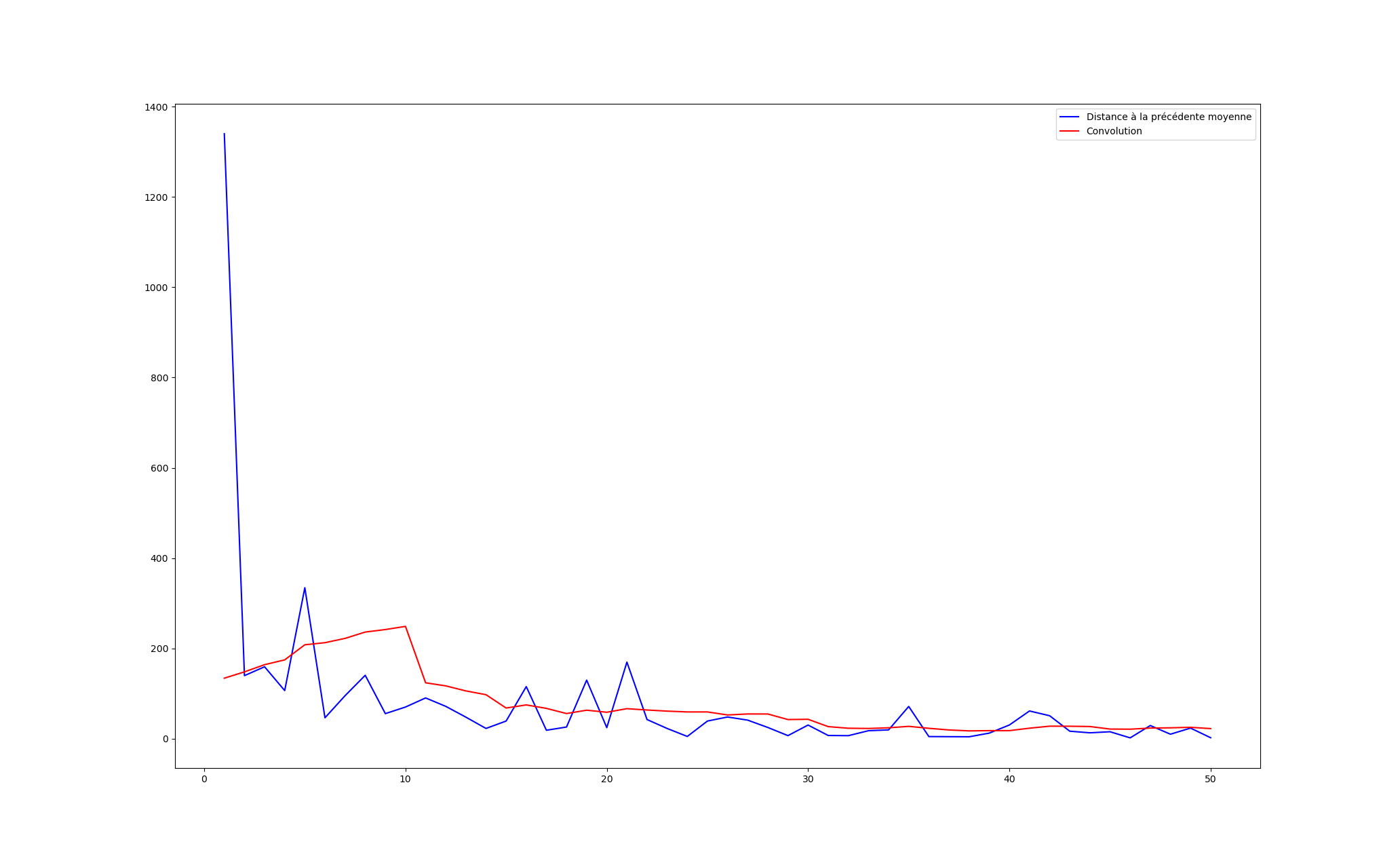}
    \caption{Studying convergence of the average} \label{fig:convmoy}
\end{figure}

We then seek to calibrate the model parameters in order to obtain results consistent with observed reality. To do this, we used a "backtesting" method. By imposing the rules of the second lockdown on the population, the parameters were varied using an exhaustive optimisation method in order to find a solution minimising the distance to the observations made thanks to the compartmental model.

\begin{figure}
    \centering
    \includegraphics[scale=0.2]{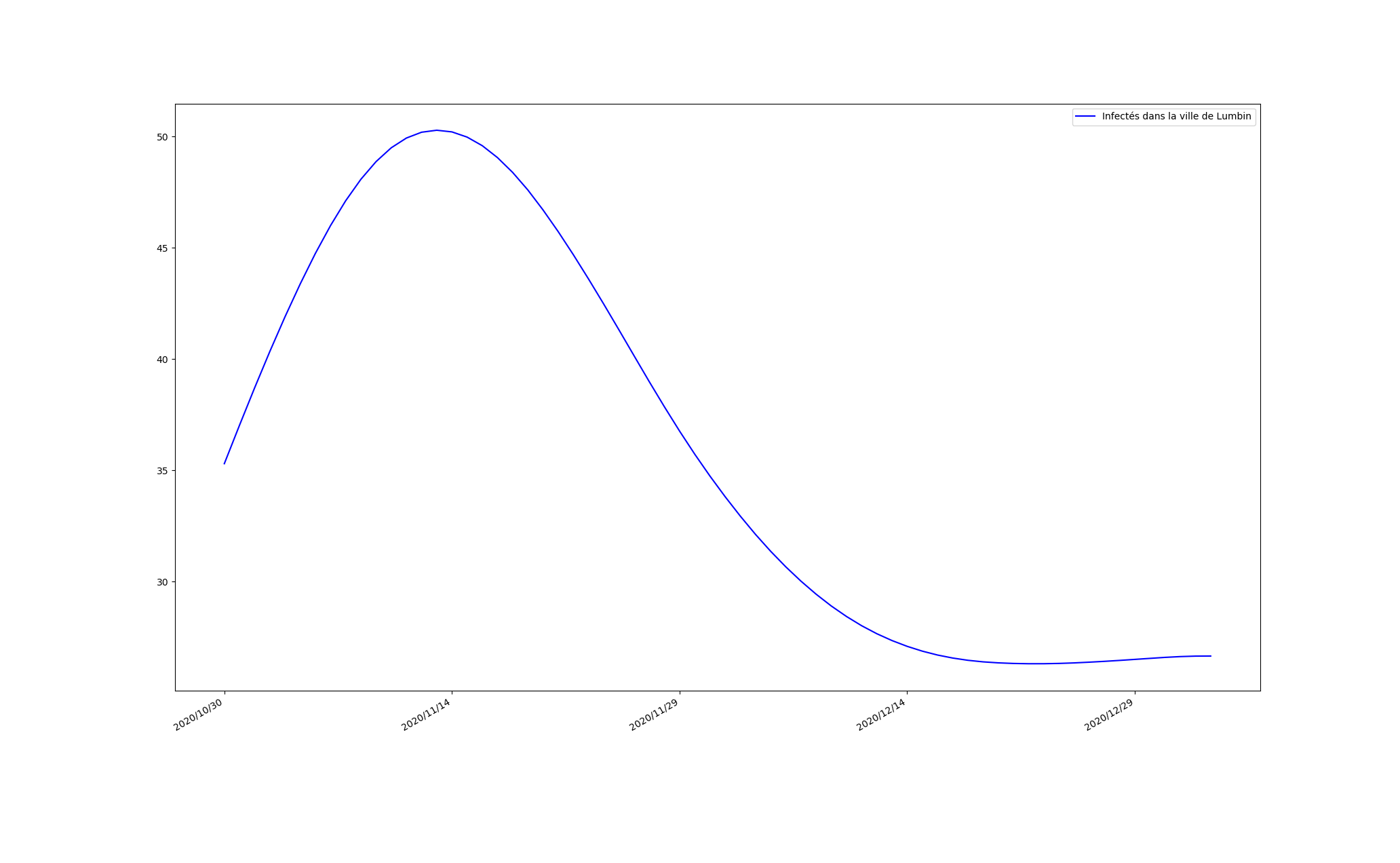}
    \caption{Number of infected people in Lumbin}
    \label{fig:LumbI}
\end{figure}

By looking at the number of cases in the town of Lumbin (see Figure~\ref{fig:LumbI}) during the second lockdown, we have an idea of what the curve should look like after the optimisation. This curve is the result of a projection of the results obtained with the Charpentier model onto the Lumbin population. It takes into account both the $I^{+}$ and $I^{-}$ compartments. It should also be noted that the lockdown ended on 15 December 2020, so this figure shows the curve that should ideally be obtained once the sanitary rules have been lifted.

\subsubsection{Results}

\begin{figure}[hbt!]
\centering
\begin{subfigure}{.45\textwidth}
    \centering
    \includegraphics[scale=0.17]{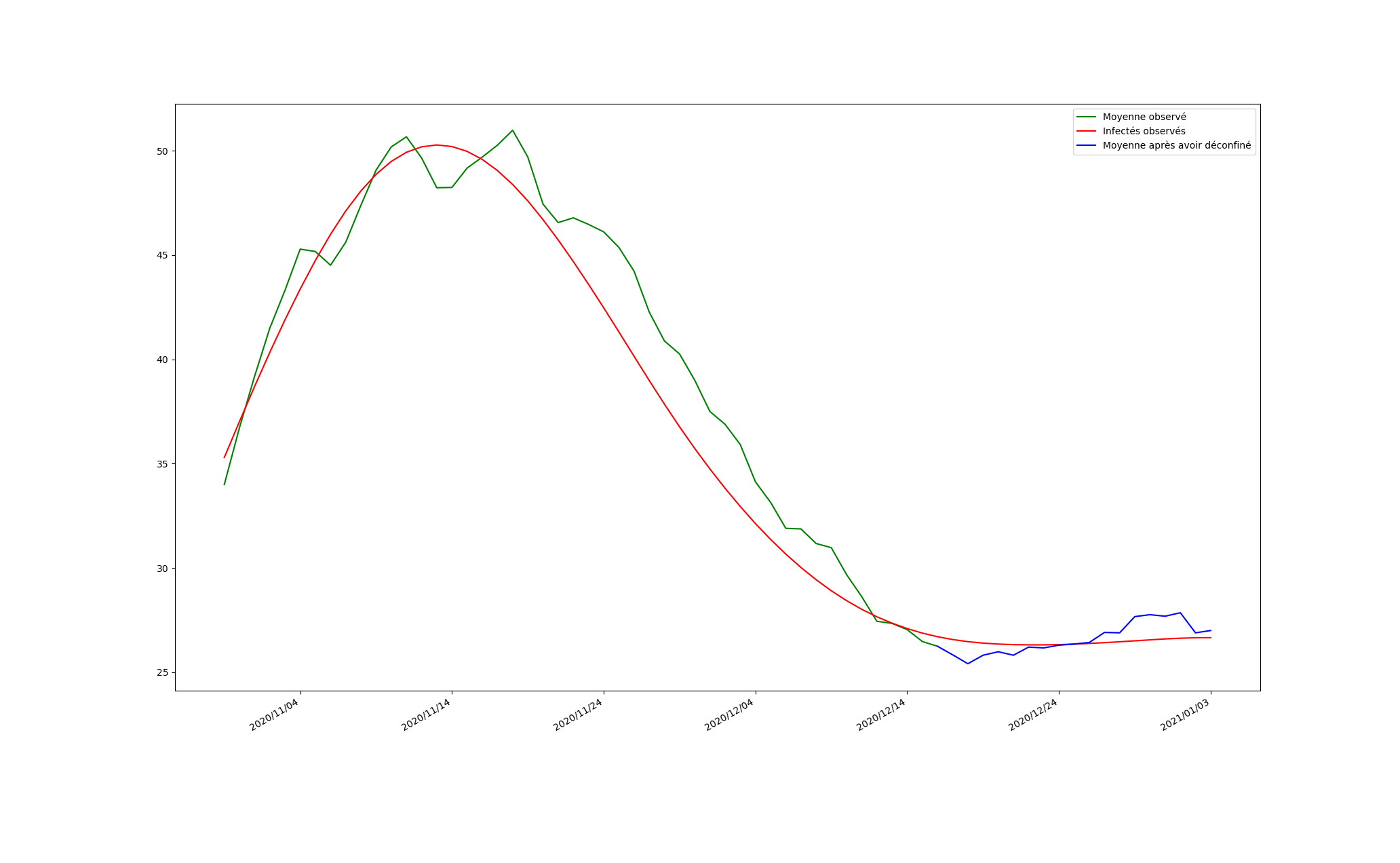}
    \caption{Results obtained after calibration}
\end{subfigure}
\begin{subfigure}{.45\textwidth}
    \centering
    \includegraphics[scale=0.17]{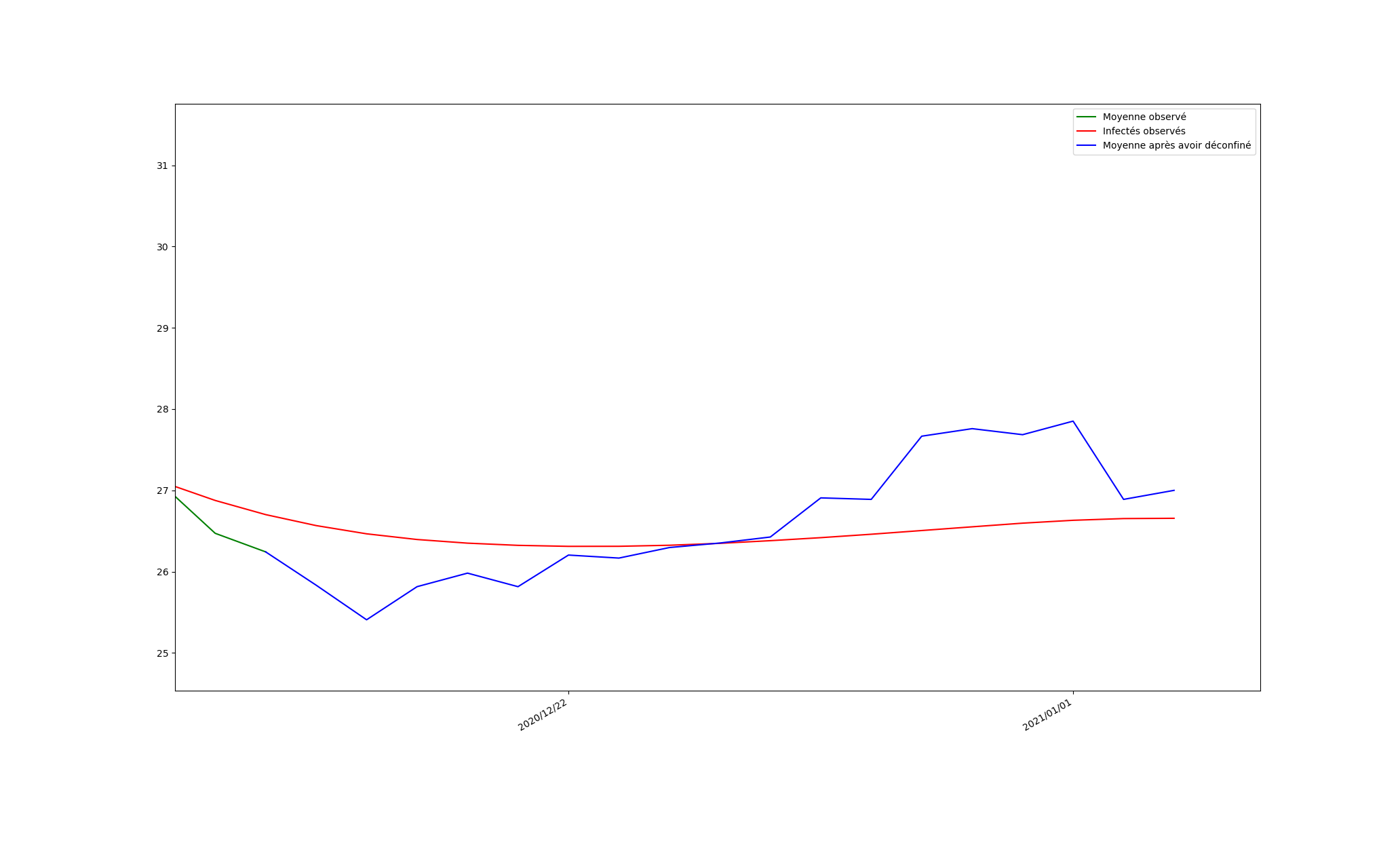}
    \caption{Zoom after lifting restrictions on 15 December 2020}
\end{subfigure}
    \caption{Number of infected estimated in the town of Lumbin: in green the calibrated part and in blue the predicted propagation after lifting restrictions}
    \label{fig:LumbExp}
\end{figure}

After calibrating the model, we obtain a multi-agent model which, by averaging the results of the experiments, comes close to the curve of the determined Infected (cf Figure~\ref{fig:LumbExp}). The green curve then represents the part calibrated in order to get closer to the observations made. On 15 December, we change the rules applied to the population: this is what we see on the blue curve. We can see that the average of the curves will stabilise at around 27 infected. We can therefore think that at a given time $t$, if we have calibrated the model on the current lockdown rules, by changing the rules in the model, we will be able to obtain a reasonable estimate of the evolution of the number of cases in the future. There are however certain points to take into account. First of all, we are studying a projection, we must keep in mind that one Lumbinois in the model represents about 600 Isérois, which means that if in the model the propagation of the error does not seem so important, in reality it will be much more important. Said differently, the less agents there are in the model, the more sensitive it is to errors. So if we apply this methodology today but we under-estimate the future propagation of the virus in Lumbin, this error will be multiplied at the scale of Isère, and future observations might widely differ from our predictions.

It is also important to note that the results presented in Figure~\ref{fig:LumbExp} all come from a single simulation. That is, the model is initialized on October 30, and the restrictions are lifted on December 15. This distinction is crucial because, due to the lockdown phase, not all age groups were impacted in the same way by the pandemic. It appears that at the end of the lockdown period, the age group least affected by the virus is the retired people, while a larger proportion of working people and children are already immune from previous exposition to the virus. If we therefore decide to simply initialise the model on 15 October by randomly distributing the number of recovered people, the simulation will be imprecise.

\section{Discussion}

\subsection{Human factors}

The work presented here allows to model the propagation of COVID-19 and how it is impacted by different policies, for instance postponing the curfew at 7pm rather than 6pm, or opening some "non-essential" shops. However, we did not evaluate the psychological impact of these policies, which is quite hard to appreciate \cite{coviprev}. Repeated lockdowns, remote classes in videoconference, lost opportunities for internships and jobs, have all had an important impact on students' life, and studies show a significant increase in the prevalence of depression in this population \cite{roux2021impact}. Healthcare workers who are in the first line are also strongly impacted \cite{caillet2021infirmiers}. Today it is obvious that mental health must be taken into account in the policies taken to fight the pandemics. 

It is also quite hard to anticipate how people might react to some measures. In particular after a crisis that has lasted for several months, tiredness starts to play a role and measures might be less respected than they were initially \cite{goldstein2021lockdown}. Risk awareness might also differ in different populations, for instance based on age or numeracy \cite{wolfe2021age}. 

Finally, the crisis leads to a loss of trust towards government and deciders \cite{oecd-trust,strandberg2020coronavirus, perry2021trust, complotiste}, and could lead to decreased trust in science \cite{zhongming2020could}. This poses major problems with the emergence of fake news \cite{enria2021trust, home2020transparency} and conspiratory theories. Using social media as a major source of information about COVID has been linked with increased mortality \cite{nieves2021infodemic}. Distrustful citizens will reject decisions, or be defiant towards the vaccine \cite{hornsey2018psychological}. 

Such individual factors as emotions, trust, or cognitive biases should therefore be taken into account in a realistic model, which is permitted by agent-based approaches \cite{bourgais2018emotion}. Future work should therefore consider the moral impact of sanitary policies, as well as their acceptability, and the evolution of trust over time as a mediator of obedience to institutional decisions \cite{vigiflood}.

\subsection{Economical and societal impact}

Economic repercussions of the sanitary crisis have contributed to widen social inequalities, whether between genders \cite{fisher2021gender} or between the rich and the poor. 

In France, it was noted that the health crisis pushed a million people into poverty \cite{1000kfr}, with temporary workers, artisans and self-employed workers being the most vulnerable categories. According to the World Bank, 150 million people throughout the world have been plunged into poverty \cite{1500k}. Again according to the World Bank, it will take 3 years to reduce the gap between the richest and the poorest if measures are taken quickly, but 10 years if nothing is done \cite{oxfam}.

The pandemic also exacerbated xenophobia. There was an increase in xenophobic acts, particularly towards people of Asian origin. Indeed, for many, they were responsible for the pandemic, but it is likely that these acts are in fact the result of systematic racism \cite{roberto2020stigmatization}. These acts are mainly found in the United States, where President Trump had presented SARS-COV-2 as the "Chinese Virus" \cite{figaroUsa}. Throughout the world, an increase in anti-Semitism has also been noted. As with xenophobia towards Asian populations, people are looking for scapegoats, and age-old stereotypes provide a designated target, particularly because of the Jewish conspiracy theory. In France, a study was carried out by the Open Society Justice Initiative \cite{justice}, which denounced the ethnic profiling carried out by the police. It was shown that during the lockdown in France, and in the framework of the application of the measures, police officers still made abusive arrests against minorities \cite{amon2020virtual}.

Such societal repercussions should be modelled to better understand how different populations react to the measures, and explain the differences in mortality.

\subsection{Environmental impact and mobility}

The COVID-19 crisis had many impacts on our environment, by changing our life mode and consumption habits. Economic activity was reduced, with a positive impact on global warming. As a result, some people believe this degrowth is beneficial and advisable even after the pandemics \cite{yildirim2022review,ramsay2020let,schaltegger2020sustainability,horisch2021relation}.

\begin{figure}[hbt!]
    \centering
    \includegraphics[scale = 0.4]{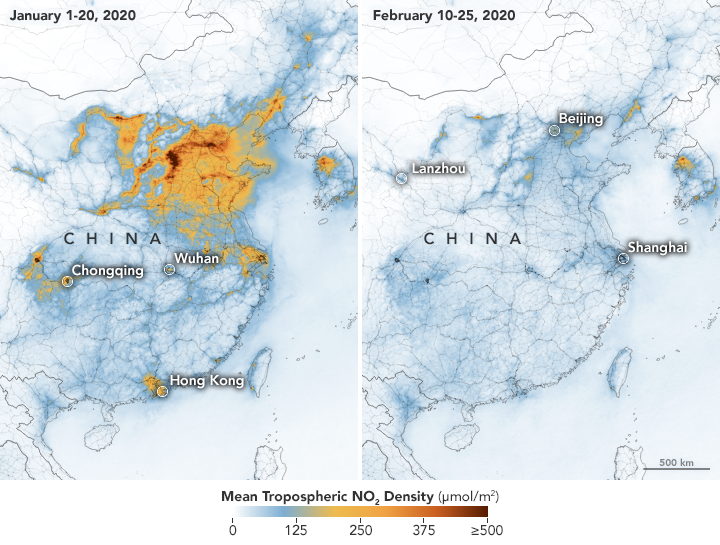}
    \caption{Nitrogen dioxide density change in China\cite{nasa-china}}
    \label{fig:china-dioxide}
\end{figure}

A first notable impact of the coronavirus epidemic is a drop in air pollution. By implementing quarantine policies and limiting travel, we have greatly reduced our carbon emissions. For example, in China, a significant drop in the density of Nitrogen Dioxide in the air was recorded over the same period between January and February 2020 \cite{carbon-china, nsabimana2020impact, nasa-china, bashir2020brief}, as seen in Figure~\ref{fig:china-dioxide}.

Improved air quality is due to reduced mobility during the lockdown. It could still be favoured by the increase in people taking their bikes for daily commuting. However it is also possible that in the future people will choose their own vehicles instead of public transport, due to raising anxiety of potential contamination in closed areas \cite{honey2020impact}.

Although the SARS-COV-2 crisis appears to have had a positive impact on the environment, there is a possibility that it could create a rebound effect. The rebound effect can be defined as "the increase in consumption linked to the reduction of limits to the use of a technology, these limits being monetary, temporal, social, physical, linked to effort, danger, organisation..." \cite{schneider2003effet}. 

A first potential for rebound effect is the digitisation of society. It allows to perform many tasks at a distance, including administrative tasks, or work (telecommuting and teleconferencing). But the expansion of Information and Communication Technologies (ICT) has a strong potential for a rebound effect \cite{freire2020pandemics, joyce2019multi}, with increased electricity  and hardware consumption.

A second potential for a rebound effect is remote working. Most companies have updated their policies about remote work, allowing people to work from home several days per week. As a result, many people have moved away from town centres \cite{nathan2020will,aaberg2021escape}, increasing the distance of their daily commute. When they do need to go to work, they might need to take their car due to a lack of public transport connection between rural and urban areas, and distances that are now infeasible by bicycle.

Simulation is a powerful tool to study such surprising or unexpected effects emerging from individual choices.

\section{Conclusion}

The work carried out here seeks to show that combining models can help beyond simple decision-making. Indeed, they make it possible to evaluate the potential impact of new sanitary measures through a projection on a town, and to extrapolate this projection to the scale of a region to measure the regional impact of these measures. However, there are some limitations. The town of Lumbin is a town of 2000 inhabitants without a cinema or a sports hall. It was therefore decided not to include these enclosed areas, where contamination by aerosol remains a possibility, which means our model is a simplification of reality. 

Several future perspectives seem interesting. First of all, it would be relevant to test the model at a larger scale, with a larger population, in order to make the results less sensitive to errors. Second, another interesting point would be to study a compartmental model taking into account the age groups with the help of intero-differential equations, as suggested by \cite{guan2020transport}. Indeed, in the current health crisis, such a study would be relevant to enable finer study of which age group is most at risk, which people to vaccinate first, or whether or not to open schools. This is interesting to evaluate finer-grained policies targeted at specific age groups, or most affected areas.

\section*{Acknowledgements}
This work was funded by the CNRS MODCOV initiative through an intership grant. This work is also related to the SEEPIA project\footnote{SEEPIA project: \url{http://51.178.55.78/SEEPIA/seepia.htm}}, the CDP RISK project\footnote{CDP RISK, Grenoble-Alps Risk institute: \url{https://risk.univ-grenoble-alpes.fr/}}, and the CovPrehension project\footnote{CovPrehension: \url{https://covprehension.org/}}. 

\footnotesize

\end{document}